# Phase Analysis on the Error Scaling of Entangled Qubits in a 53-Qubit System

Wei-Jia Huang[1], Wei-Chen Chien[2], Chien-Hung Cho[1], Che-Chun Huang[1], Tsung-Wei Huang [3], Seng Ghee Tan[4], Chenfeng Cao[5], Bei Zeng[5], and Ching-Ray Chang[2,6,7]

[1]Department of Physics, National Taiwan University, Taipei, Taiwan
[2]Graduate Institute of Applied Physics, National Taiwan University, Taipei, Taiwan
[3]Department of information and computer Engineering, Chung Yuan Christian University, Taiwan
[4]Department of Optoelectric Physics, Chinese Culture University, 55 Hwa-Kang Road, Yang-Ming-Shan, Taipei 11114, Taiwan
[5]Department of Physics, The Hong Kong University of Science and Technology
[6]Graduate Institute of Electronics Engineering, National Taiwan University, Taipei, Taiwan
[7]NTU-IBM Quantum Hub, National Taiwan University

E-mail: r06222062@ntu.edu.tw

We have studied carefully the behaviors of entangled qubits on the IBM Rochester with various connectivities and under a "noisy" environment. A phase trajectory analysis based on our measurements of the GHZ-like states is performed. Our results point to an important fact that entangled qubits are "protected" against environmental noise by a scaling property that impacts only the weighting of their amplitudes. The reproducibility of most measurements has been confirmed within a reasonably short gate operation time. But there still are a few combinations of qubits that show significant entanglement evolution in the form of transitions between quantum states. The phase trajectory of an entangled evolution, and the impact of the sudden death of GHZ-like states and the revival of newly excited states are analyzed in details. All observed trajectories of entangled qubits arise under the influences of the newly excited states in a "noisy" intermediate-scale quantum (NISQ) computer.

## 1. Introduction

Quantum entanglement[1-3] is an important index of a truly observable quantum phenomenon. This phenomenon occurs when the nonlocality of a pair of particles is generated due to mutual interactions. Therefore, the independent quantum state of each particle cannot be relied upon to understand the physical phenomenon of entanglement. Even when the pair is separated over a long distance, quantum entanglement might still persist. This is the major difference between classical and quantum physics. Quantum entanglement has been the focus of intense theoretical and experimental research[4] for its potentially wide applications. The existence of quantum entanglement and hence the applicability of the $2^N$ Hilbert space[5,6] are the major advantages of quantum computers compared to classical. Many specific indicators, such as quantum volume[7] and Mermin's inequality[8], provide the theoretical quantification to determine whether quantum entanglement exists in a multi-qubit system.

Quantum entanglement and decoherence are closely related. Natural physical systems are usually not completely isolated from the external world and the result of interactions with the environment is the major source of decoherence. According to quantum mechanics, entanglement creates associations between the constituent quantum states of subsystems. Quantum nonlocality is generally described as equivalent to entanglement. It is also considered as a requisite condition for quantum teleportation[9,10] and quantum cryptography.[11-13] Many experiments have already demonstrated that electrons, photons, neutrinos, molecules and even diamond vacancies show quantum entanglement.[14,15] The use of entanglement in communication,[16] computing[17] and quantum radar[18] is a very active area of research and development. From previous reported results[19] for the IBM Rochester[20], entanglement of a large number of qubits are easily affected by the environmental noise, but the entanglement states of a small number of qubits are relatively stable[19]. Also, our measurements provide another easy test for the entanglement of a N-qubit system[19]. Generally, a 2-qubit pair uses Bell's[21] or Mermin's inequality[8] to distinguish its "quantum-ness" from local realism (LR). The GHZ-like states[22-24] are used as the initial states for both Bell's and Mermin's. Although the phase angles tested for the maximum values of LR are different, the basic physics for both are similar – difference lies only in the superposition of the subsystem quantum states. For multiple qubits, maximal values of Mermin's polynomials are often relied upon to understand the entanglement physics of a N-qubit system and its quantum subsystems[19]. However, transitions between states can occur if the energy levels of the NISQ[25] system fluctuate (Supplement A). In some cases, entanglement can disappear completely within a finite interval - this is sometimes known as the "Entangled Sudden Death" (ESD)[26]. On the other hand, decoherence is related to the relaxation time $T_1$ and the dephasing time $T_2$ [27]. This phenomenon has recently attracted the interest of many researchers because it directly affects the dynamic performance of a quantum computer. Actual implementation of quantum computation and quantum communication depends on the lifetime of the qubits. One of the most difficult obstacles that must be overcome for the development of fault-tolerant quantum computers is to fully understand the evolution mechanisms of the entangled states. Entangled states might lose coherence due to interaction with the environments. Therefore, entangled states collapse because of the aforementioned de-coherence, as well as due to the process of measurement. In order to successfully develop a fault-tolerant universal quantum computer, it is necessary to have a full understanding of the evolution of entangled states and the procedures of quantum measurement. In the following, we would carefully study the behavior of the two-qubit subsystems on the IBM Rochester under a noisy environment.

Phase analysis[28-30] is a common method to study responses in classical systems. The initial phase used in the GHZ-like state[22-24] in Mermin's polynomials has a natural advantage for the phase trajectory analysis of an entangled system - in particular the evolution of transitions between states. Taking the GHZ-like state as initial, we apply the phase trajectory analysis on the IBM Rochester and systematically explore the effects of various initial phase angles. Measurement results would depend on the measurement period, cycle and time. We explore the evolution of entangled pairs with different initial phase angles. Several patterns of phase trajectory are observed in our measurements. Our phase trajectory analysis shows "normal" and "abnormal" circles of amplitude variation, and that superposition can transit between states (Supplement A). The evolution of the superposition of entangled states with noise are also studied (Supplement C) based on noise models discussed[31,32]. Most quantum measurements of phase trajectory on the IBM Rochester give repeatable circles but with different radius. However, some specific combinations of qubits (e.g. connection 4-6) are very unstable and irreproducible even run within a very short time interval. Phase trajectory analysis shows interesting entanglement evolution, evident in the varying shapes of the so-called abnormal circles. Some entanglement evolution even switches between large and small circles, and the radius of entanglement exceeds the LR limit. From the Mermin's polynomials' point of view[8], the entanglement of a 2-qubit state can suddenly disappear (i.e., within LR value) and revive at a later stage (i.e., outside of LR value). Our quantum computer measurement results are further compared with numerical analysis with noise (Supplement C) as well as classical simulations (Supplement D). It can be concluded that all observed trajectories arise due to the entanglement properties of the newly excited states in a NISQ computer.

## 2. Methods and Theory

The two qubits Mermin's polynomials[8,22] are
$$\begin{cases} M_2 = XX + XY + YX - YY \\ M_2' = -XX + XY + YX + YY \end{cases} (1)$$
Usually the GHZ-like state[22-24] is used to measure the Mermin's inequalities and the 2-qubit GHZ-like state[22-24] is
$$|GHZ_2\rangle = \frac{1}{\sqrt{2}}(|00\rangle + e^{i\varphi}|11\rangle) \ (2)$$
The advantage of GHZ-like initial state is that phase analysis can be easily implemented for measurements[19], with the variation of $e^{i\varphi}$. Also maximum value can be obtained for Mermin's polynomials at a certain phase angle.

The expectation values of the Mermin's polynomials for a 2-qubit are easily derived and
$$\begin{cases} \langle M_2 \rangle = 2\sqrt{2}\cos\left(\varphi - \frac{1}{4}\pi\right) \\ \langle M_2' \rangle = 2\sqrt{2}\sin\left(\varphi - \frac{1}{4}\pi\right) \end{cases} (3)$$
Here we would modify the Mermin's polynomials[8] and actually carry out measurements on IBM Rochester with $\langle W_2 \rangle$ and its associated $\langle W_2' \rangle$.

$$\begin{cases} W_2 = XX + 2YX - YY \\ W_2' = -XX + 2YX + YY \end{cases} (4)$$

For a pure GHZ-like state[22-24], since $\langle XY \rangle + \langle YX \rangle = 2\langle YX \rangle$, measurements for $\langle W_2 \rangle \langle W_2' \rangle$ as opposed to $\langle M_2 \rangle \langle M_2' \rangle$ give exactly the same values. However, because of the environmental noise in a NISQ system, there are other possible states besides the initial GHZ-like states, that can be excited before measurement (Supplement A). A possible excited entangled state (Supplement A) can be written as

$$\rho' = \begin{pmatrix} 0 & 0 & 0 & 0 \\ 0 & a & re^{-i\theta} & 0 \\ 0 & re^{i\theta} & 1-a & 0 \\ 0 & 0 & 0 & 0 \end{pmatrix} (5)$$

which is a quantum state in the subspace spanned by $|01\rangle$ and $|10\rangle$, $r$, $a$, $\theta$ are parameters that determines the density matrix of $\rho'$. The newly entangled state $\rho'$ can be generated as a result of energy fluctuation, and

$$Tr(\rho' M_2) = Tr(\rho' M'_2) = 0 \ (6)$$

Therefore, one could not observe noise-excited entanglement states in a NISQ system with the conventional $\langle M_2 \rangle$ and $\langle M_2' \rangle$. However, with our modified measurements of $\langle W_2 \rangle$ and $\langle W_2' \rangle$,

$$Tr(\rho' W_2) = Tr(\rho' W'_2) = 4r\sin\theta \quad (7)$$

as shown in Eq. (7), the entangled evolution of noise-excited states in a NISQ system can be easily measured. In other words, $\langle W_2 \rangle$ and $\langle W'_2 \rangle$ measurements allow the study of phase trajectory portraits of not only the GHZ-like states, but also any noise-induced quantum states. Therefore, the modified $\langle W_2 \rangle$ and $\langle W'_2 \rangle$ measurements will be used for our phase trajectory analysis throughout.

## 2.1 Quantum circuit

A single qubit with a *H* gate can produce a GHZ-like state, and the second qubit is entangled with the GHZ-like state. A $U1(\varphi)$ gate then operates on the qubit from the *H* gate and together the three gates form a complete quantum circuit in our oracle. For a 2-qubit entanglement testing, we set $|GHZ_2\rangle = \frac{1}{\sqrt{2}}(|00\rangle + e^{i\varphi}|11\rangle)$ as shown in Fig. 1(a).

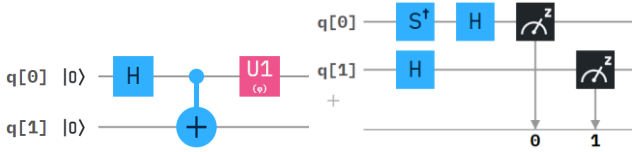

Fig. 1: (a) The quantum circuit for a 2-qubit GHZ-like state, $|GHZ_2\rangle = \frac{1}{\sqrt{2}}(|00\rangle + e^{i\varphi}|11\rangle)$. *H* represents the Hadamard gate, *U1* (φ) is a gate that rotates a quantum state about the z axis to impart phase φ, and *CNOT* gate entangles the two qubits. (b) The quantum circuit of *YX measurement for the 2-qubit pair*.

## 3. Results

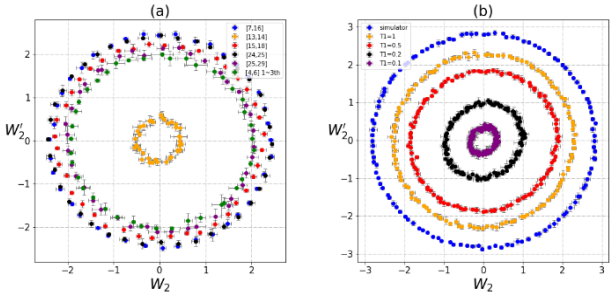

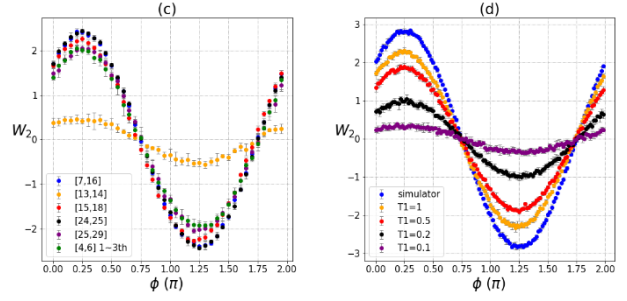

Figure 2： Phase trajectories of $\langle W_2 \rangle$ and $\langle W'_2 \rangle$ (a and b) and the relationships of $\langle W_2 \rangle$ with the initial phase angles φ (c and d). Measurements are carried out for 2-qubit pairs on the IBM Rochester. Experimental shots are 1,024; each data point is the average of five measurements; and the average values and standard variations are plotted. (a) Phase trajectories for six different 2-qubit pairs with different initial phase angles φ. (b) Classical simulation of 2-qubit pairs with different $T_1$. (c) $\langle W_2 \rangle$ and initial phase angle *φ*. (d) Classical simulation results for different $T_1$.

We have carried out phase trajectory analysis for 2-qubit systems on the IBM Rochester and also simulations on classical computer with different relaxation time $T_1$ (Fig. 2). Figure 2a shows the measurements of $\langle W_2 \rangle$ and $\langle W'_2 \rangle$ on the IBM Rochester. Figure 2b shows the classical simulation of $\langle W_2 \rangle$ and $\langle W'_2 \rangle$ for different $T_1$. All 2-qubit pairs on the IBM Rochester were measured[19], but we only showed results for chosen pairs of [7,16], [13,14], [15,18], [24,25], [25,29]. In particular, we carried out three different measurements for pair [4,6]. Figure 2a clearly shows that the entangled radii are different for all different 2-qubit pairs, even though they could be of the same phase. Measurement results were then compared with classical simulations involving different $T_1$. From Fig. 2b, it is clear that radii of the circles shrink as the $T_1$ value decreases. In a NISQ computer, the performance of qubits is commonly affected by the noise environments. In the classical simulations, we used the Qiskit module with gate time equal to 0.1 second, and different $T_1$ are determined through the fitting of the amplitude of $|1\rangle$ (Supplement D). Larger $T_1$ means less environment noise and the coherence of quantum entanglement sustains for a longer interval. From classical simulations, the amplitude of the superposition of the initial states will be affected by the environment noise. But the circular phase trajectories seem impervious to the environment noise as long as states $|00\rangle$ and $|11\rangle$ remain.

To study the noise, we introduced parameters $\gamma_0 t$ and $\gamma_1 t$ to the amplitude of the GHZ-like state and the noise excited $\rho'$ (Supplement A). Considering both $|\Psi\rangle$ and the newly excited $\rho'$ on a NISQ computer, measurements of $\langle W_2 \rangle$ and $\langle W'_2 \rangle$ show different trajectories arising due to the influence of the entanglement strength (characterized by parameter $\gamma_0 t$ and $\gamma_1 t$ ). Our theoretical analytic results and classical simulations of the influences of $|00\rangle$, $\rho'$ and $|11\rangle$ are given in Supplement C and D. Quantum measurements on the IBM NISQ computer indeed testifies strongly to our theory of amplitude transition between states, evident in the significant changes of the circular radii. The superposition amplitude of the GHZ-like states are indeed very sensitive to the environment, but the persistence of the circular trajectories speaks for an important fact, i.e. their entanglement is reasonably robust. In fact, the redistribution of the amplitudes suggests that the noise induces energy transition between states $|00\rangle$ and $|11\rangle$.

It is important to note that the circular shapes persist (Fig. 2a) and the sinusoidal waves are all in phase (Fig. 2c). Classical simulations show that the amplitudes of sinusoidal waves will decrease as $T_1$ decreases (Fig. 2c) but no phase shift is induced by any environmental noise. This indicates that entangled qubits are "protected" against environmental noise by a scaling property that impacts only the weighting of their amplitudes (Supplement C).

Besides the commonly circular, some unusual trajectories were also observed in our measurements for $\langle W_2 \rangle$ and $\langle W_2' \rangle$. For qubit pair [4,6], we observed a lot of peculiar patterns from December 2019 to January 2020. Measurements on other entangled pairs are mostly repeatable, but pair [4,6] could not be reproduced over different cycles. It should be noted that these observations only existed for pair [4,6] during that period of time. We had carried out seven measurements for pair [4,6]; patterns of trajectories are shown below and the underlying mechanisms are discussed.

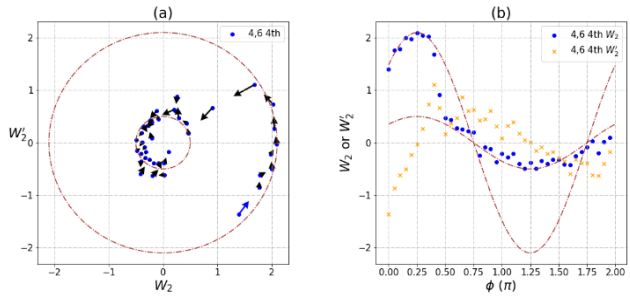

Figure 3: A phase trajectory of transition pattern between small and large circles for $\langle W_2 \rangle$ and $\langle W_2' \rangle$ measurements of pair [4,6] at 4th run on IBM Rochester. Experimental shots are 1024. (a) The trajectory of $\langle W_2 \rangle$ and $\langle W_2' \rangle$. (b) The relationship between $\langle W_2 \rangle$, $\langle W_2' \rangle$ and phase angle φ. The radius of outer circle is 2.1 and inner circle is 0.5. The maximal value of GHZ-like state is $2\sqrt{2}$.

There are two circular trajectories shown in Fig. 3, and a direct transition from the large circle with radius 2.1 to the small circle with radius 0.5 is observed. Large and small circles are still in phase but with different amplitudes (Fig. 3b). The change of radius for the circular trajectories indicates that the quantum system is dissipative, and the amplitude weight for state $|11\rangle$ reduces accordingly. From Fig. 2, transitions from large to small circles can be linked to the reduction of $T_1$, and it is thus clear that the abrupt change of radius (Fig.3a) is the result of a sudden increase in the environment noise at that point.

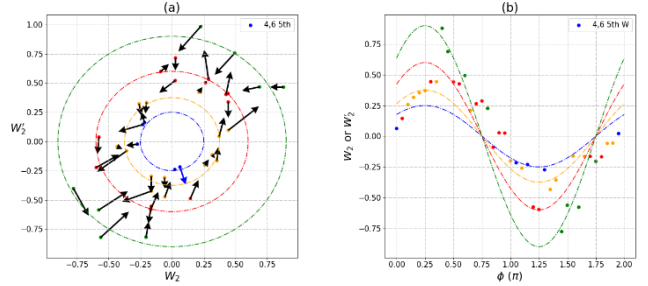

Fig 4: A transition pattern between four circles for $\langle W_2 \rangle$ and $\langle W_2' \rangle$ measurement of pair [4,6]. (a) The trajectories of $\langle W_2 \rangle$ and $\langle W_2' \rangle$. (b) The relationship between $\langle W_2 \rangle$, $\langle W_2' \rangle$ and phase angles φ. Data measured on the IBM Rochester are marked in four different colors to denote the different radii. Transition between circles are fluctuating in the NISQ system. The maximal value of the GHZ-like state is $2\sqrt{2}$.

Environmental noise is the source of instability responsible for all the amplitude fluctuation in a NISQ computer. This leads to the transitions between four circles that we observed in Fig. 4. However, data of the same color can still fit the circle of a specific radius. Since the circles still persist, entanglement between $|00\rangle$ and $|11\rangle$ remains robust at all times (Supplement C). Once again environment noise fails to destroy the entanglement but does impact heavily on the weights of the superposition amplitudes. Here, the four radii of approximately 0.25, 0.375, 0.6, 0.9 represent their normalized maximal values of the Mermin's polynomial.

However, it is worth noting that measurement results did show phase shifts in certain cases. This is in spite of the fact that the input phase of the GHZ-like states were initially assigned in all those cases. Some unexpected backward dots could also be found in this measurement result. The straightforward speculation here is that large noise environment could impact both the entangled phase and amplitude.

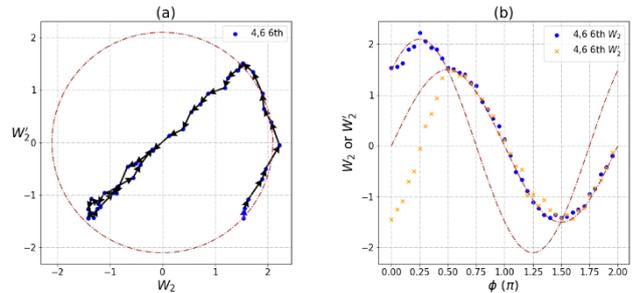

Fig 5: A transition pattern between a circle and a line for $\langle W_2 \rangle$

and $\langle W_2'\rangle$ of pair [4,6]. (a) The trajectory of $\langle W_2\rangle$ and $\langle W_2'\rangle$. (b) The relationship between $\langle W_2\rangle$, $\langle W_2'\rangle$ and phase angles φ. The radius of circle is 2.1. The maximal value of the GHZ-like state is $2\sqrt{2}$.

Other than transitions between several circles of different radii, an interesting phase trajectory that shows up in line and circular paths is observed in Fig. 5. It is obvious that the initially partial circle comes from the superposition of states $|00\rangle$ and $|11\rangle$. From Eq. (7), if state $|11\rangle$ dissipate into ρ′, *i.e.,* a state supported by $|01\rangle$ and $|10\rangle$ (see supplement C), a linear projection based on the measurement of $\langle W_2\rangle$ and $\langle W_2'\rangle$ should be observed. The GHZ–like state will be mixed with state ρ′ which arises as an excitation from the noisy environments during the dynamic processes of entanglement. Even though the initial states are the GHZ-like states only, states $|00\rangle, |01\rangle, |10\rangle, |11\rangle$ can combine in a superposition fashion in a NISQ computer. From our numerical simulations and analytic analysis (Supplement C,D), the line along the $\langle W_2\rangle=\langle W_2'\rangle$ direction can be explained by the vanishing of the high energy entangled state $|11\rangle$. The graph in Fig. 5 clearly demonstrates a transition of the higher energy $|11\rangle$ to state ρ′. However, if the conventional $\langle M_2\rangle$ and $\langle M_2'\rangle$ were adopted for measurements instead, the trajectory of the entangled states would have shrunk to a point instead of showing up as a line. This makes the study of their entanglement behaviors much more difficult. The benefit of using the $\langle W_2\rangle$ and $\langle W_2'\rangle$ measurements becomes obvious here. From Eqs. (3) to (7), the phase dependence of the $\langle W_2\rangle$ measurement for the GHZ-like state is $cos\left(\varphi-\frac{\pi}{4}\right)$, while that for a general excited state ρ′, where $\rho' = \begin{pmatrix} 0 & 0 & 0 & 0 \\ 0 & a & re^{-i\theta} & 0 \\ 0 & re^{i\theta} & 1-a & 0 \\ 0 & 0 & 0 & 0 \end{pmatrix}$, $\langle W_2\rangle$ measurement for the GHZ-like state is $4r\sin(\theta)$. For some noise models, such as the depolarizing noise located before the CNOT gate (Supplement C), $\sin(\theta) = \sin(\varphi)=\cos(\varphi-\frac{\pi}{2})$. The $|11\rangle$ state "dies" suddenly and states ρ′ appear at the crossing point of the circular and the line paths. A π/4 phase shift observed in Fig. 5b clearly supports the interpretation around the sudden death of state $|11\rangle$ and the sudden birth of states ρ′.

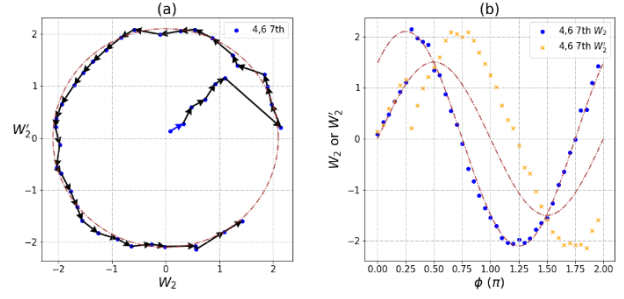

Fig 6: A phase trajectory showing a transition from lines to the circular paths for $\langle W_2\rangle$ and $\langle W_2'\rangle$ measurement of pair [4,6]. (a) The trajectory of $\langle W_2\rangle$ and $\langle W_2'\rangle$. (b) The relationship between $\langle W_2\rangle$, $\langle W_2'\rangle$ and phase angles φ. The radius of circle is 2.1. The maximal value of the GHZ-like state is $2\sqrt{2}$.

Even though the GHZ-like state is prepared for measurement, the initial line trajectory in Fig. 6 indicates that state $|11\rangle$ disappears immediately after it is assigned. Instead, state ρ′ is very much alive since the beginning. However, the rebirth of state $|11\rangle$ changes the line back into the large circular path again. This result suggests that energy transfer between the IBM Rochester qubits and the environment (i.e., circuitry and control system to interact with the qubits) is indeed fluctuating throughout the course of many measurements. Phase shift around φ = π/4 in Fig. 6(b) also implies the sudden emergence of state ρ′.

## 4.Discussion and Conclusion

Today's quantum computer is still pretty much a NISQ system. Many research efforts are now focused on the elimination of noises[33], and the emulation of quantum computers on classical platforms[34]. Applications the likes of quantum adiabatic optimization algorithms[35], variational quantum eigensolvers,[36] hash preimage attacks[37], and modeling of viral diffusion,[38] are to all still run on NISQ computers. Therefore, understanding the phase trajectory of measurements for entangled qubits will speed up their eventual adoptions on quantum computers[39]. The entanglement of pair [4,6] produces trajectories beyond circular path, which suggests the effect of energy fluctuation in a NISQ system is significant for certain connectivity of qubits. For an energy stable NISQ, the phase trajectory is always constant, i.e., the trajectory is always circular, while for unstable and noisy quantum computer, multiple circular paths of different radii can be observed. The observation of line trajectories from both simulation and experimental results on the IBM Rochester can also be understood from our noise and classical simulations (supplement C and D). The missing of state $|11\rangle$ due to insufficient energy on the quantum computer is the underlying cause for the interesting phase trajectory observed in our analysis. From Fig. 5 and 6, a large circular path becomes a line along direction $\langle W_2\rangle=\langle W_2'\rangle$ and vice versa. This observation shows that the system energy fluctuates heavily and the sudden death and birth of quantum states occur all the time. In other words, entanglement strength and superposition of entangled states do constantly fluctuate in a noisy environment. Nonetheless, entanglement for other qubit pairs persists in all measurements, only the amplitudes of their superposition states vary. This constancy of the radii during phase analysis suggested the scaling possibility of error mitigation for different chosen qubits connectivity.

In summary, we have developed a modified Mermin's polynomials, and applied them to study the phase trajectory of quantum entanglement on a IBM Q 53-qubit quantum computer. Most of the qubit pair results fall within the prediction of the Mermin's polynomials. Pair [4,6] shows a very strange behavior though and did not exactly obey the LR predictions of Bell's inequality and Mermin's polynomials. The observation of a large circular path with radius outside of the LR limit confirmed its state of entanglement. But a small circle, within the LR limit, that still shows quantum correlations of measurement, cannot be explained by the hidden variable or the physics of LR. In particular, a straight line along the diagonal direction is also observed within the LR limit, and this suggests that the quantum states die a sudden death under a noisy environment but revive again later. We use both classical simulations and theoretical analysis to study our measurement results from the IBM Rochester. We conjectured that the line trajectories within the LR limit could still be a result of entanglement. The projection of Hilbert's space onto the classical world gets modified with our use of the modified Mermin's polynomials of $\langle W_2 \rangle$ and $\langle W_2' \rangle$. Although straight lines are not the typical results, they could still represent entanglement. In fact, a π/4 phase shift in Figs. 5(b) and 6(b) at the line-circle path crossing point supports the existence of entanglement. We offer a more plausible explanation from the point of view of quantum entanglement of states $|01\rangle$ and $|10\rangle$. Sudden deaths and revivals of quantum states do not destroy entanglement. They merely show a lack of energy in the system to sustain the higher energy state. Therefore, we can conclude that the IBM Rochester shows a reasonably good performance with entangled qubits even for the very unstable pair [4,6]. The phase trajectory within LR is a projection of quantum entangled states subject to the fluctuation of system energy. Last, the NISQ IBM Rochester does still sport unstable qubit pairs, e.g. [4,6] which should be avoided by users.

**Acknowledgements**

Thanks for NTU-IBM Q Hub at National Taiwan University from Ministry of Science and Technology, Taiwan, under grant No. MOST 107-2627-E-002-001-MY3, 108-2627-E-002-002 and NTU-107 L104064 provide quantum computer resources.

# Supplementary Information
## Supplement A

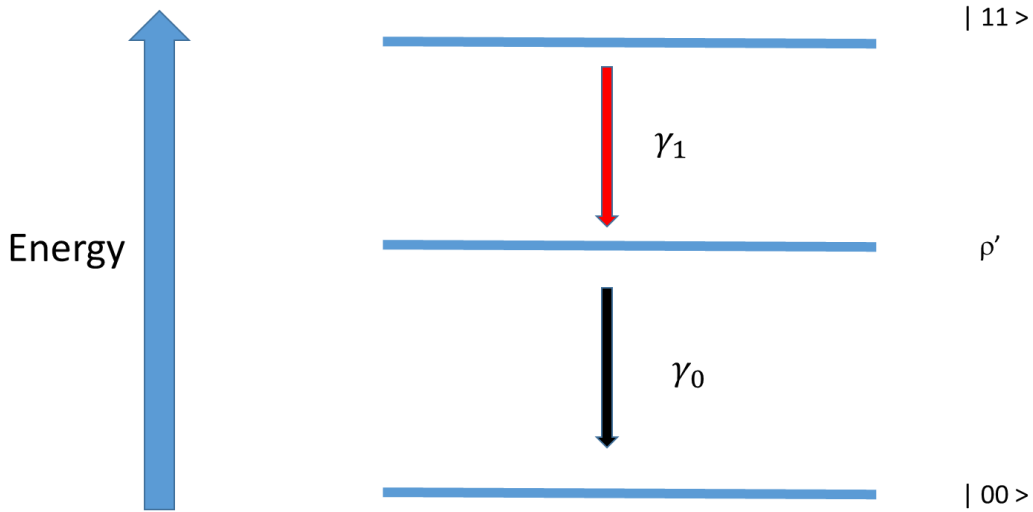

Figure SA: Four states |11>, |01>, |10> and |00> of a 2-qubit pair. $\rho'$ is a quantum state in the subspace spanned by $|01\rangle$ and $|10\rangle$, which is usually mixed. It can be written as

$$\rho' = \begin{pmatrix} 0 & 0 & 0 & 0 \\ 0 & a & re^{-i\theta} & 0 \\ 0 & re^{i\theta} & 1-a & 0 \\ 0 & 0 & 0 & 0 \end{pmatrix},$$

Possible transitions between states are shown for an energy dissipative system, and the ground state |00> is assumed to be always alive. $\gamma_1$ represents the transition rate from |11> to $\rho'$, while $\gamma_0$ is from $\rho'$ to |00>.

## Supplement B

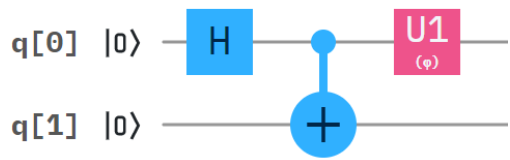

Figure SB: Operating on q[0] ⊗ q[1] with the Hadamard gate and the CNOT gate.

After operating on q[0] ⊗ q[1] with the Hadamard gate and the CNOT gate, possible transitions in a noisy environment are shown in Fig. SB. The resultant state will usually be not $\frac{1}{\sqrt{2}}(|00\rangle + |11\rangle)$, the state first assigned to the quantum computer. The noise-induced transitions generate q[0] ⊗ q[1] = $A|00> +B|01> +C|10> +D|11>$. However, operator $U1(\varphi) = \begin{pmatrix} 1 & 0 \\ 0 & e^{i\varphi} \end{pmatrix}$ acting on this state will impart a phase of $e^{i\varphi}$ to state |1> in q[0]. This usually results in the final state of $A|00> +B|01> +Ce^{i\varphi}|10> +De^{i\varphi}|11>$.

## Supplement C

In this section, we present a method with uncorrelated errors. We know if the circuit shown in Fig. 1 is noiseless, the corresponding circle in Fig. 2 would be the largest circle with radius $\sqrt{|\langle\Psi|W_2|\Psi\rangle|^2 + |\langle\Psi|W_2'|\Psi\rangle|^2} = 2\sqrt{2}$. The circuit is accompanied by uncorrelated noise, as shown in Fig. SC1, where we have considered models for three different noise sources: the depolarizing channel, the dephasing channel and the amplitude damping channel [31,32].

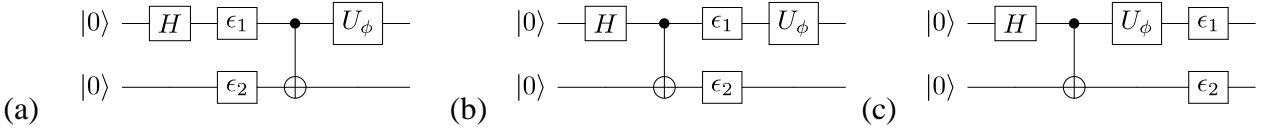

Figure SC1: Noisy Circuit with noise channels $\varepsilon_1$ and $\varepsilon_2$. (a) Noise before the CNOT gate. (b) Noise after the CNOT gate. (c) Noise after the phase rotation gate.

Denote the noise rate as *p*, the depolarizing noise channel is modeled by

$$\rho \to (1-p)\rho + pI/2$$

The dephasing noise is modeled by

$$\rho \to \sum_i E_i \rho E_i^\dagger$$

where $E_0 = \begin{pmatrix} \sqrt{1-p} & 0 \\ 0 & \sqrt{1-p} \end{pmatrix}$, $E_1 = \begin{pmatrix} \sqrt{p} & 0 \\ 0 & 0 \end{pmatrix}$, $E_2 = \begin{pmatrix} 0 & 0 \\ 0 & \sqrt{p} \end{pmatrix}$. The amplitude damping noise is modeled by

$$\rho \to A_0 \rho A_0^\dagger + A_1 \rho A_1^\dagger$$

where $A_0 = \begin{pmatrix} 1 & 0 \\ 0 & \sqrt{1-p} \end{pmatrix}$, $A_1 = \begin{pmatrix} 0 & \sqrt{p} \\ 0 & 0 \end{pmatrix}$. Suppose ρ is the density matrix of the state generated from the noisy circuit, $p_1$ and $p_2$ are the noise rates of channels $\varepsilon_1$ and $\varepsilon_2$, respectively.

When the noise channel is located prior to the CNOT gate,

the density matrix for the depolarizing noise is given by

$$\rho = \begin{pmatrix} \frac{1}{2}-\frac{1}{4}p_2 & 0 & 0 & (\frac{1}{2}-\frac{1}{2}p_1-\frac{1}{4}p_2+\frac{1}{4}p_1p_2)e^{-i\varphi} \\ 0 & \frac{1}{4}p_2 & (\frac{1}{4}p_2-\frac{1}{4}p_1p_2)e^{-i\varphi} & 0 \\ 0 & (\frac{1}{4}p_2-\frac{1}{4}p_1p_2)e^{i\varphi} & \frac{1}{4}p_2 & 0 \\ (\frac{1}{2}-\frac{1}{2}p_1-\frac{1}{4}p_2+\frac{1}{4}p_1p_2)e^{i\varphi} & 0 & 0 & \frac{1}{2}-\frac{1}{4}p_2 \end{pmatrix}$$

Notice that here the excited state $\rho' = \begin{pmatrix} 0 & 0 & 0 & 0 \\ 0 & a & re^{-i\theta} & 0 \\ 0 & re^{i\theta} & 1-a & 0 \\ 0 & 0 & 0 & 0 \end{pmatrix}$ has parameters $a = \frac{1}{2}, r = \frac{1}{2} - \frac{1}{2}p_1, \theta = \varphi$, which means $\rho' = \begin{pmatrix} 0 & 0 & 0 & 0 \\ 0 & \frac{1}{2} & (\frac{1}{2}-\frac{1}{2}p_1)e^{-i\varphi} & 0 \\ 0 & (\frac{1}{2}-\frac{1}{2}p_1)e^{i\varphi} & \frac{1}{2} & 0 \\ 0 & 0 & 0 & 0 \end{pmatrix}$.

A special case is if $p_1 = 0$, then $\rho' = |\Psi'\rangle\langle\Psi'|$ with $|\Psi'\rangle = \frac{1}{\sqrt{2}}(|01\rangle + e^{i\varphi}|10\rangle)$.

The density matrix for the dephasing noise is

$$\rho = \begin{pmatrix} \frac{1}{2} & 0 & 0 & (\frac{1}{2}-\frac{1}{2}p_1)e^{-i\varphi} \\ 0 & 0 & 0 & 0 \\ 0 & 0 & 0 & 0 \\ (\frac{1}{2}-\frac{1}{2}p_1)e^{i\varphi} & 0 & 0 & \frac{1}{2} \end{pmatrix}$$

Last, the density matrix for the amplitude damping noise is

$$\rho = \begin{pmatrix} \frac{1}{2}+\frac{1}{2}p_1 & 0 & 0 & \frac{1}{2}\sqrt{1-p_1}e^{-i\varphi} \\ 0 & 0 & 0 & 0 \\ 0 & 0 & 0 & 0 \\ \frac{1}{2}\sqrt{1-p_1}e^{i\varphi} & 0 & 0 & \frac{1}{2}-\frac{1}{2}p_1 \end{pmatrix}$$

When the noise channel is located after the CNOT gate or the phase rotation gate, the density matrix for the depolarizing noise is given by

$$\rho = \begin{pmatrix} \frac{1}{2}-\frac{1}{4}p_1-\frac{1}{4}p_2+\frac{1}{4}p_1p_2 & 0 & 0 & (\frac{1}{2}-\frac{1}{2}p_1-\frac{1}{2}p_2+\frac{1}{2}p_1p_2)e^{-i\varphi} \\ 0 & \frac{1}{4}p_1+\frac{1}{4}p_2-\frac{1}{4}p_1p_2 & 0 & 0 \\ 0 & 0 & \frac{1}{4}p_1+\frac{1}{4}p_2-\frac{1}{4}p_1p_2 & 0 \\ (\frac{1}{2}-\frac{1}{2}p_1-\frac{1}{2}p_2+\frac{1}{2}p_1p_2)e^{i\varphi} & 0 & 0 & \frac{1}{2}-\frac{1}{4}p_1-\frac{1}{4}p_2+\frac{1}{4}p_1p_2 \end{pmatrix}$$

Notice that here for the excited state $\rho'$, $a = \frac{1}{2}$, $r = 0$. $\rho' = \begin{pmatrix} 0 & 0 & 0 & 0 \\ 0 & \frac{1}{2} & 0 & 0 \\ 0 & 0 & \frac{1}{2} & 0 \\ 0 & 0 & 0 & 0 \end{pmatrix}$.

The density matrix for the dephasing noise is given by

$$\rho = \begin{pmatrix} \frac{1}{2} & 0 & 0 & (\frac{1}{2}-\frac{1}{2}p_1-\frac{1}{2}p_2+\frac{1}{2}p_1p_2)e^{-i\varphi} \\ 0 & 0 & 0 & 0 \\ 0 & 0 & 0 & 0 \\ (\frac{1}{2}-\frac{1}{2}p_1-\frac{1}{2}p_2+\frac{1}{2}p_1p_2)e^{i\varphi} & 0 & 0 & \frac{1}{2} \end{pmatrix}$$

Last, the density matrix for the amplitude damping noise is

$$\rho = \begin{pmatrix} \frac{1}{2}+\frac{1}{2}p_1p_2 & 0 & 0 & \frac{1}{2}\sqrt{1-p_1-p_2+p_1p_2}\,e^{-i\varphi} \\ 0 & \frac{1}{2}p_1-\frac{1}{2}p_1p_2 & 0 & 0 \\ 0 & 0 & \frac{1}{2}p_2-\frac{1}{2}p_1p_2 & 0 \\ \frac{1}{2}\sqrt{1-p_1-p_2+p_1p_2}\,e^{i\varphi} & 0 & 0 & \frac{1}{2}-\frac{1}{2}p_1-\frac{1}{2}p_2+\frac{1}{2}p_1p_2 \end{pmatrix}$$

Notice that here the parameters for $\rho'$ are $a = \frac{\frac{1}{2}p_1-\frac{1}{2}p_1p_2}{\frac{1}{2}p_1+\frac{1}{2}p_2-p_1p_2}$, $r = 0$.

We conclude that if there is depolarizing noise before CNOT gate (which is usually true in experiments), $\theta = \varphi$.

Therefore, we can easily calculate the relation between radius $R = \sqrt{tr^2(\rho W_2) + tr^2(\rho W_2')}$ and noise rates $p_1, p_2$. For example, when the amplitude damping channel is located after the CNOT gate, $R =$ $2\sqrt{2(1-p_1-p_2+p_1p_2)}$. In addition to the depolarizing noise, in the case of noise before the CNOT gate, R does not depend on the phase φ, therefore the phase trajectory is always a circle with radius R for different noise channels and noise rates. If we assume $p = p_1 = p_2$, the radii for the different noise channels and noise rates will be as shown in Fig. SC2.

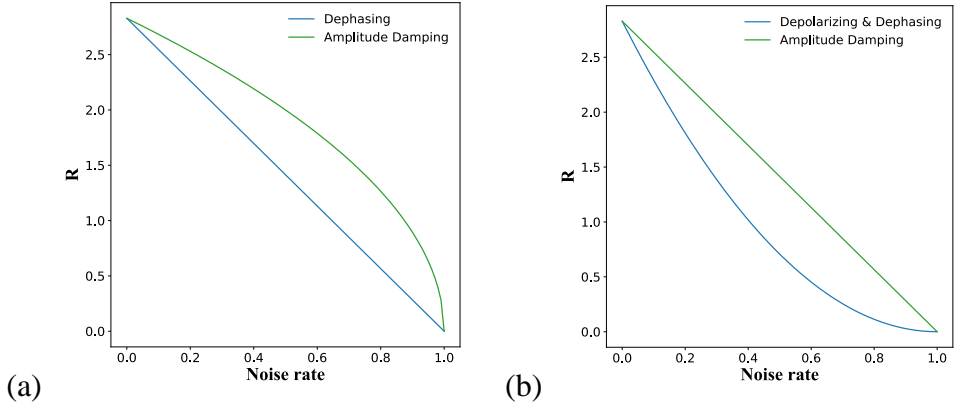

(a)  (b)

Fig SC2: Radii for different noise channels and noise rates. (a) Result for system of noise channel located before the CNOT gate. (b) Result for system of noise channel located after the CNOT gate or after the phase rotation gate.

If the amplitude damping channel is located after the CNOT or the phase rotation gate, we will have $p = 1 - \frac{R}{2\sqrt{2}}$. It is also known that noise rate is $p = 1 - e^{-\frac{t}{T_1}}$ for the amplitude damping noise. From the two equations above, we have $\ln\left(\frac{2\sqrt{2}}{R}\right) = \frac{t}{T_1}$, which is consistent with the simulation results in Fig.2(b). The trajectories of the noise rates in Fig. SC3 correspond to the trajectories in Fig. 2(b) with different $T_1$.

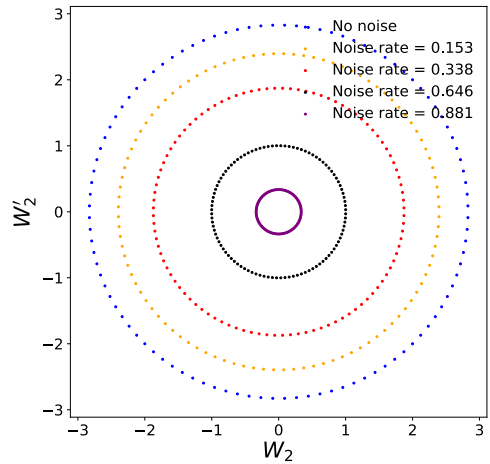

Fig SC3: Phase trajectories for the uncorrelated amplitude damping channels located after the CNOT/phase gate.

When the depolarizing channel is located before the CNOT gate, the trajectory will be ellipses instead of circles, as shown in Fig SC4.

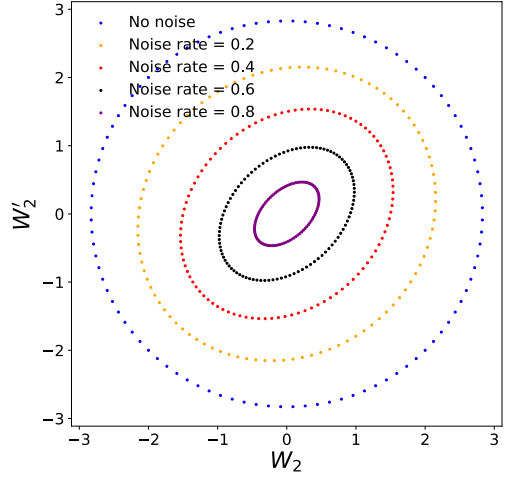

Fig SC4: Phase trajectories for the uncorrelated depolarizing channel is located before CNOT gate.

We will now also take $T_2$ into consideration. If the noise channel is a combination of the amplitude damping noise and the dephasing noise, it will transform a single qubit density matrix through

$$\rho = \begin{pmatrix} 1 - \rho_{11} & \rho_{01} \\ \rho_{01}^* & \rho_{11} \end{pmatrix} \to \begin{pmatrix} 1 - \rho_{11} e^{-\frac{t}{T_1}} & \rho_{01} e^{-\frac{t}{T_2}} \\ \rho_{01}^* e^{-\frac{t}{T_2}} & \rho_{11} e^{-\frac{t}{T_1}} \end{pmatrix}.$$

For a 2-qubit state $|\Psi\rangle\langle\Psi|$, where $|\Psi\rangle = \frac{1}{\sqrt{2}}(|00\rangle + e^{i\varphi}|11\rangle)$, we denote $T_1^1$ and $T_2^1$ as the relaxation time $T_1$ and dephasing time $T_2$ of the first qubit, $T_1^2$ and $T_2^2$ as, respectively, the "times" for the second qubit. The combined noise channel transforms $|\Psi\rangle\langle\Psi|$ to

$$\rho = \begin{pmatrix} 1 - \frac{1}{2}e^{-\frac{t}{T_1^1}} - \frac{1}{2}e^{-\frac{t}{T_1^2}} + \frac{1}{2}e^{-\frac{t}{T_1^1}-\frac{t}{T_1^2}} & 0 & 0 & \frac{1}{2}e^{-\frac{t}{T_2^1}-\frac{t}{T_2^2}}e^{-i\varphi} \\ 0 & \frac{1}{2}e^{-\frac{t}{T_1^2}} - \frac{1}{2}e^{-\frac{t}{T_1^1}-\frac{t}{T_1^2}} & 0 & 0 \\ 0 & 0 & \frac{1}{2}e^{-\frac{t}{T_1^1}} - \frac{1}{2}e^{-\frac{t}{T_1^1}-\frac{t}{T_1^2}} & 0 \\ \frac{1}{2}e^{-\frac{t}{T_2^1}-\frac{t}{T_2^2}}e^{i\varphi} & 0 & 0 & \frac{1}{2}e^{-\frac{t}{T_1^1}-\frac{t}{T_1^2}} \end{pmatrix}$$

The phase trajectories will still be circles.

# Supplement D

## Classical simulation results of phase angle and orthogonal measurements for state $|\Psi'\rangle$

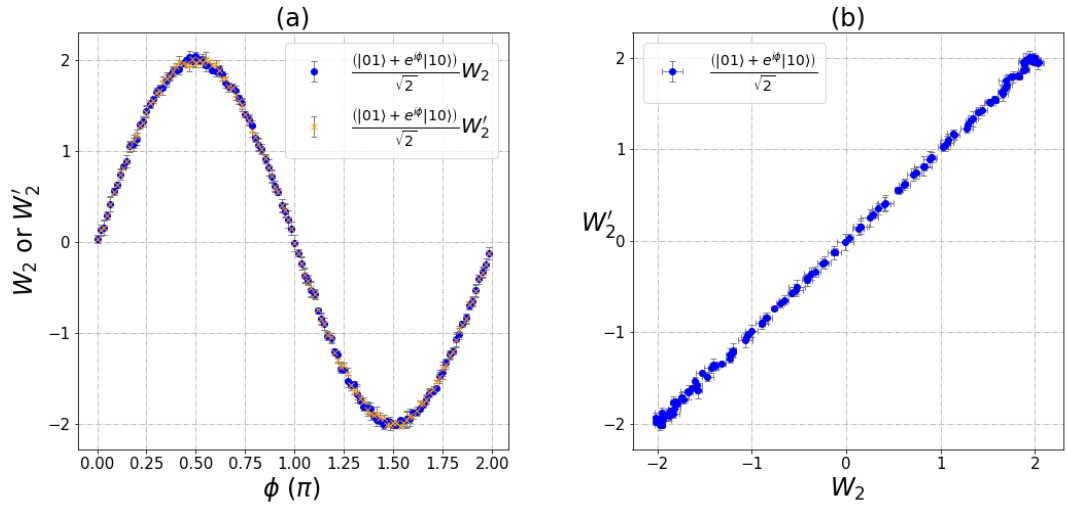

Fig SD: The classical simulation results of superposition states with $|\Psi'\rangle = \frac{1}{\sqrt{2}}(|01\rangle + e^{i\varphi}|10\rangle)$. (a) The relationship of $\langle W_2 \rangle$, $\langle W_2' \rangle$ with phase angle $\varphi$. (b) The relationship between $\langle W_2 \rangle$ and $\langle W_2' \rangle$.